\documentclass[aps,prl,twocolumn,superscriptaddress,showpacs,reprint]{revtex4}

\usepackage[dvips]{graphicx}
\usepackage{amsmath}

\begin{document}

\title{Avalanche Collapse of Interdependent Networks}

\author{G. J. Baxter}
\email[]{gjbaxter@ua.pt}
\affiliation{Department of Physics \& I3N, University of Aveiro, Campus
  Universit\'ario de Santiago, 3810-193 Aveiro, Portugal}
\author{S. N. Dorogovtsev}
\author{A. V. Goltsev}
\affiliation{Department of Physics \& I3N, University of Aveiro, Campus
  Universit\'ario de Santiago, 3810-193 Aveiro, Portugal}
\affiliation{A. F. Ioffe Physico-Technical Institute, 194021
  St. Petersburg, Russia}
\author{J. F. F. Mendes}
\affiliation{Department of Physics \& I3N, University of Aveiro, Campus
  Universit\'ario de Santiago, 3810-193 Aveiro, Portugal}

\date{\today}

\begin{abstract}
We reveal the nature of the avalanche collapse of the giant viable
component in multiplex networks under perturbations such as random
damage.
Specifically, we identify latent critical
clusters
associated with the avalanches of random damage. Divergence of their mean
size signals the approach to the hybrid phase transition from one
side, while there are no critical precursors on the other side. We
find that this discontinuous transition occurs in scale-free multiplex
networks whenever the mean degree of at least one of the
interdependent networks does not diverge.
\end{abstract}

%\keywords{}

\maketitle

%=====================================================================
Many complex systems, both natural
\cite{Pocock2012}, and man-made \cite{Rinaldi2001,Kurant2006}, can be
represented as multiplex or interdependent networks.
Multiple dependencies make a system
more fragile: damage
to one element can lead to avalanches of failures throughout the system
\cite{Osorio2007,Poljansek2012}.
Recent theoretical investigation of two  \cite{Buldyrev2010} or more
\cite{Gao2011} networks
in which vertices in each network mutually depend on vertices in
other networks has shown that indeed small initial failures
can cascade
back and forth through the networks, leading to a discontinuous
collapse of the whole system.
Damage in one network propagates along edges and leads to damage in
the other network. This is an individual stage of a cascade in
back-and-forth damage propagation.
Son {\it et al.}. \cite{Son2012} showed that this approach can be
simplified and is equivalent to considering damage propagation in
multiplex networks. They proposed a simple mapping between
the model used in \cite{Buldyrev2010} 
in which a vertex in one network
has a mutual dependence on exactly one vertex in the other network,
and a multiplex network with one kind of vertex but two kinds of
edges. The mapping is achieved by simply merging the mutually dependent
vertices from the two networks.

\begin{figure}[ht]
\includegraphics[width=1.0\linewidth]{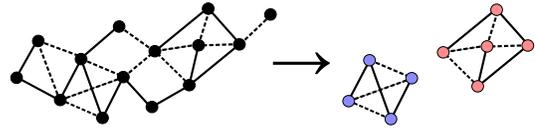}
\caption{A small
  network with two kinds of edges %, dashed and solid
(left).
Applying
  the algorithm described in the text non-viable vertices are
  removed leaving two viable clusters (right).}\label{viable_cluster}
\end{figure}

In this Letter we describe the nature of such discontinuous phase transitions.
We consider a set of vertices connected by $m$ different types of
edges (dependencies).
The connections are essential to the function of each site, so that a
vertex is only viable if it maintains connections of every type to
other viable vertices. A \emph{viable cluster} is defined as follows:
For every kind of edge, and for any two
vertices $i$ and $j$ within a viable cluster, there must be a path from $i$
to $j$ following only edges of that kind. A graph containing two
finite viable
clusters is illustrated in Fig. \ref{viable_cluster}.
We wish to find when there is a giant cluster of viable vertices.
Note that any giant viable cluster is a subgraph
of the giant connected component of each of the $m$ networks.

%=================================================================
%================================================================

\begin{figure}[ht]
\includegraphics[width=1.0\linewidth]{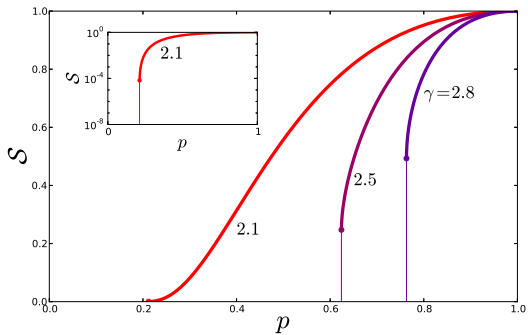}
\caption{Size of the giant viable cluster $\mathcal{S}$ as a function
  of the fraction $p$ of vertices remaining undamaged for two
  symmetric powerlaw distributed networks with, from
  right to left, $\gamma = 2.8$ , $2.5$, and $2.1$. The height of the jump
  becomes very small as $\gamma$ approaches $2$, but is not zero, as
  seen in the inset, which is $\mathcal{S}$ vs $p$ on a
  logarithmic vertical scale for $\gamma=2.1$.}\label{S_vs_p}
\end{figure}

Various parameters can be used to control the critical
behavior of this system:
 the mean degrees of the networks, amount of random damage and so on.
Small perturbations to the system can propagate, leading to avalanches
of further damage.
In uncorrelated, random networks
 we find a discontinuous
hybrid transition in the collapse of the giant viable cluster, similar
to that seen in the $k$-core or bootstrap percolation
\cite{Dorogovtsev2006a,BDG2}.
In Ref. \cite{Buldyrev2010}, the propagation of damage caused by
removal of a finite fraction of vertices in one of the interdependent
networks was studied. In contrast, we study the avalanches
of damage triggered by the removal of randomly chosen single vertices.
These avalanches
increase in size approaching the critical point, signaling the
impending collapse of the giant viable cluster. At the critical point
the mean avalanche size diverges.
Below the transition, on the other hand, there is no precursor for the
appearance of the giant viable cluster.
The transition is thus asymmetric. It is hybrid in nature, having a
discontinuity like a
first-order transition, but exhibiting critical behavior, only above
the transition, like a second-order transition. A complete
understanding of the transition cannot therefore be had without first
understanding this critical behavior.
We have discovered critical clusters
which collapse in avalanches of diverging size as the transition is
approached. These critical clusters are thus responsible for both the
critical scaling and the discontinuity observed in the size of the
giant viable cluster.
As we shall see, the critical clusters have a novel
character as, unlike the corona clusters of the $k$-core for example
\cite{Dorogovtsev2006a},
avalanches propagate in a directed way through critical clusters.
The critical clusters may have important practical applications,
helping to identify vulnerabilities to targeted attack,
as well as informing efforts to guard against such attack.
Surprisingly, when the degree distributions are asymptotically
power-law $P(q) \propto
q^{-\gamma}$ the critical point $p_c$ (taking the undamaged fraction of
vertices $p$ as the control parameter)
 remains at a finite
value even when the exponents $\gamma$ of the degree distributions are
below $3$, remaining finite until both exponents reach $2$, in
agreement with an argument given in \cite{Buldyrev2010}. This is in
stark contrast to ordinary percolation in complex networks, in which
the threshold falls to zero as soon as $\gamma$ reaches $3$
\cite{Albert2000,Callaway2000}.
We show, further, that the nature of the transition doesn't
change. Although the height of the discontinuity becomes extremely
small near $\gamma=2$, it remains
finite near this limit (see Fig.~\ref{S_vs_p}).

%=================================================================
%=================================================================

{\it Algorithm}.---We consider a multiplex network, with vertices $i =
1, 2, ..., N$
connected by $m$ kinds of edges labeled $s = a, b, ...$.
The joint degree distribution is $P(q_a,q_b,...)$.
Viable clusters in any multiplex network may be identified by the
following algorithm.\\
\noindent(i)
Choose a test vertex $i$ at random from the network.\\
(ii)
For each kind of edge $s$, compile a list of vertices that can be
reached from $i$ by following only edges of type $s$.\\
(iii)
The intersection of these $m$ lists forms a new candidate set for
the viable cluster containing $i$.\\
(iv)
Repeat steps (ii) and (iii) but traversing only the current candidate
set. When the candidate set no longer changes, it is either a
viable cluster, or contains only vertex $i$.\\
(v)
To find further viable clusters, remove the viable cluster of $i$
from the network (cutting any edges) and repeat steps (i)-(iv) on the
remaining network beginning from a new test vertex.

Repeated application of this procedure will identify every viable
cluster in the network. The application of this procedure to a finite
graph is illustrated in Fig.~\ref{viable_cluster}.

%=================================================================
\begin{figure}[ht]
\includegraphics[width=1.0\linewidth]{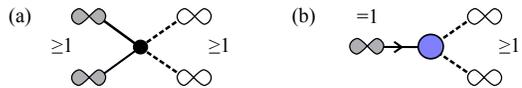}
\caption{Viable and critical viable vertices for two interdependent
  networks.
(a) A vertex is in the giant viable cluster if it
  has connections to giant viable subtrees (represented by infinity
  symbols) of both kinds.
(b) A critical viable vertex of type $a$ has exactly $1$
  connection to a giant sub-tree of type $a$.
}\label{active_and_corona}
\end{figure}
%================================================================

%=================================================================
\begin{figure}[ht]
\includegraphics[width=1.0\linewidth]{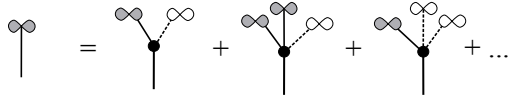}
\caption{Diagrammatic representation of Eq. (\ref{Psi_s}) in a system
  of two interdependent networks $a$ and $b$. The probability $X_a$,
  represented by a shaded infinity symbol
can be written recursively as a sum of second-neighbor
  probabilities. Open infinity symbols represent the equivalent
  probability $X_b$ for network $b$, which obeys a similar recursive
  equation.
}\label{Xa_diagram}
\end{figure}
%================================================================

{\it Basic Equations}.---Let us consider the case of sparse
uncorrelated
networks, which are locally tree-like in the infinite size limit $N
\to \infty$. In such a network there are no finite viable clusters.
In order to find the giant viable cluster,
we define $X_s$, with $s \in \{a,b,...\}$,
 to be the probability that, on following an
arbitrarily chosen edge of type $s$, we encounter the root of an
infinite sub-tree formed solely from type $s$ edges, whose vertices
are also each connected to at least one infinite subtree of every
other type. We call this a type $s$ infinite subtree.
The vector $\{X_a,X_b,...\}$ plays the role of
the order parameter.
A vertex is then in the giant viable cluster if it has at least
one edge of every type $s$ leading to an infinite type $s$ sub-tree
(probability $X_s$), as shown in Fig.~\ref{active_and_corona}(a).
Using the locally tree-like property of the networks, we can write
self consistency equations for the probabilities $X_s$:
\begin{multline}\label{Psi_s}
X_s = \Psi_s(X_a,X_b,...)\equiv\\
\sum_{q_a,q_b,...}\!\!\!\!\frac{q_s}{\langle q_s\rangle}
P(q_a,q_b,...) \big[1-(1-X_s)^{q_s-1} \big] \!\prod_{l\neq s}\!
\big[1-(1-X_l)^{q_l}\big]
\end{multline}
for each $s \in \{a,b,...\}$. This is illustrated in
Fig.~\ref{Xa_diagram}.
The term $({q_s}/{\langle q_s\rangle}) P(q_a,q_b,...)$ gives the
probability that on following an arbitrary edge of type $s$, we
find a vertex with degrees $q_a, q_b,...$, while
$[1-(1\!-\!X_a)^{q_a}]$ is the probability that this vertex has at
least one edge of type $a \neq s$ leading to the root of an
infinite sub-tree of type $a$ edges (i.e. probability $X_a$). This
becomes $[1-(1\!-\!X_s)^{q_s-1}]$ when $a = s$.
The argument leading to Eq. (\ref{Psi_s}) is similar to that used
in \cite{Son2012}.
Solving these equations enables us to calculate all the quantities of
interest. In particular, the relative size of the giant viable cluster
is
\begin{equation}\label{S}
\mathcal{S} = \sum_{q_a,q_b,...}\!\! P(q_a,q_b,...)\!\prod_{s=a,b,...}\!\!
\big[1-(1\!-\!X_s)^{q_s} \big],
\end{equation}
which is illustrated in Fig.~\ref{active_and_corona}(a).

A hybrid transition appears at the point where $\Psi_s(X_a,X_b,...)$ first
meets $X_s$ at a non-zero value, for all $s$.
This occurs when
\begin{equation}\label{hybrid_condition_asym}
\det[{\bf J}-{\bf I}] = 0
\end{equation}
 where ${\bf I}$ is the unit matrix and
${\bf J}$ is the Jacobian matrix
$J_{ab} = \partial \Psi_b/\partial X_a$.
Expanding $\Psi_s$ about the critical point, at which Eqs. (\ref{Psi_s})
and (\ref{hybrid_condition_asym}) are both satisfied, we find
the scaling of %that
$X_s$ and hence $\mathcal{S}$, the
size of the giant viable cluster.
For example, random damage can be considered by introducing a
parameter $p$, the fraction of vertices remaining undamaged. This is
incorporated by multiplying the right-hand sides of Eqs. (\ref{Psi_s})
and (\ref{S}) by a factor $p$.
 Then
\begin{equation}\label{squareroot}
\mathcal{S} - \mathcal{S}_c \propto X_s - X_s^{(c)} \propto (p-p_c)^{1/2}.
\end{equation}
A similar result is found for other control parameters.

\begin{figure}[ht]
\includegraphics[width=1.0\linewidth]{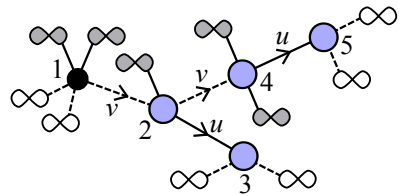}
\caption{
A critical cluster. Removal of any of the shown viable vertices will
result in the removal of all downstream critical viable
vertices. Removal of the vertex labeled $1$
will result in all of the shown vertices being removed (becoming
non-viable), while removal of vertex $4$ results only in vertex $5$
also being removed.
}\label{critical_cluster}
\end{figure}

{\it Avalanches}.---To examine the hybrid transition we focus on the
case of two types of edges.
Consider a viable vertex that has exactly one edge of type $a$ leading
to a type $a$ infinite subtree, and at least one edge of type $b$
leading type $b$ infinite subtrees. We call this a
critical vertex of type $a$. It is illustrated in
Fig.~\ref{active_and_corona}(b). 
Critical vertices of type $a$ will
drop out of the viable cluster if they lose their single link to a
type $a$ infinite subtree.
We mark these special edges with an arrow leading to the critical
vertex.
An avalanche can only transmit in the direction of the arrows.
A vertex may have outgoing edges of this kind, so that removal of this
vertex from the giant viable cluster also requires the removal of the
critical vertices which depend on it. For example, in
Fig.~\ref{critical_cluster}, removal of the vertex labeled 1 removes
the essential edge of the critical vertex 2 which thus becomes
non-viable. Removed critical viable vertices
may in turn have outgoing critical edges, so that the removal of a
single vertex can result in an avalanche of removals of critical vertices
from the giant viable cluster.
In Fig.~\ref{critical_cluster}, removal of 2
causes the removal of further critical vertices 3 and 4, and the
removal of 4 then requires the removal of 5.
Thus critical vertices form critical clusters. At the head of each
critical cluster is a `keystone vertex' (e.g. vertex $1$ in the
figure)  whose removal would result in
the removal of the entire cluster.
Graphically, upon removal of a vertex,
we remove all vertices found by following the arrowed
edges.
As we approach the critical point (from above), 
 diverging
avalanches cause a discontinuity in the size of the
 giant viable cluster, which collapses to zero.

%=================================================================
%=================================================================

There are three possibilities when following an
arbitrarily chosen edge of a given type: i) with probability $X_s$ we
encounter a type $s$ infinite subtree  ii) with probability
$R_s$ we encounter a vertex which has a connection to an infinite
subtree of the opposite type, but none of the same type. Such a vertex
is part of the giant viable cluster if the parent vertex was; or
iii) with probability $1-X_s-R_s$, we encounter a vertex which has no
connections to infinite subtrees of either kind.
The probability $R_a$ obeys
\begin{equation}
R_a = \sum_{q_a}\sum_{q_b}\! \frac{q_a}{\langle q_a\rangle}P(q_a,q_b)
(1\!-\!X_a)^{q_a-1} \left[1\! -\! (1\!-\!X_b)^{q_b}\right]
\label{R1}
\end{equation}
 and similarly for $R_b$.
We use generating functions to examine the sizes of critical
clusters.
We first define the function $F_a(x,y)$ as
\begin{align}
F_{a}(x,y) =& \sum_{q_a}\sum_{q_b}\! \frac{q_a}{\langle
  q_a\rangle}P(q_a,q_b) x^{q_a-1}\sum_{r=1}^{q_b}\!\binom{q_b}{r} X_b^r
y^{q_b-r}
\label{F1_asym}
\end{align}
and similarly for $F_{b}(x,y)$, by exchanging all subscripts $a$ and
$b$.
The generating function for the size of 
an avalanche triggered by removing
an arbitrary type $a$ edge which does not
lead to an infinite type $a$ subtree can be defined in terms of these
functions by
\begin{equation}\label{H1a_asym}
H_{a}(u,v) = 1-X_a-R_a + uF_a[H_{a}(u,v),H_{b}(u,v)]
\end{equation}
and similarly for $H_{b}(u,v)$.
This recursive equation can be understood by noting
that $H_{a}(0,v) = 1-X_a-R_a$ is the probability that an
arbitrarily chosen edge leads to a vertex outside the
viable cluster.
Here $u$ and $v$ are auxiliary variables. Following through a critical
cluster, a factor $u$ appears for each arrowed edge of type $a$, and
$v$ for each arrowed edge of type $b$. For example, the critical
cluster illustrated in Fig.~\ref{critical_cluster} contributes a
factor $u^2 v^2$.
The mean number of critical vertices reached upon
following an edge of type $a$, i.e. the mean size of the resulting
avalanche if this edge is removed, is given by
$\partial_u H_a(1,1)+\partial_v H_a(1,1)$.
Unbounded avalanches emerge at the point where $\partial_u H_a(1,1)$
[or $\partial_v H_{b}(1,1)$] diverges.
Taking derivatives of Eq. (\ref{H1a_asym}),
 and using that $H_{a}(1,1) = 1\!-\!X_a$ and $F_{a}(1\!-\!X_a,1\!-\!X_b)
= R_a$,
and that, from Eqs. (\ref{Psi_s}) and (\ref{F1_asym}),
$\partial_x F_{a}(1\!-\!X_a,1\!-\!X_b) = \partial_a \Psi_a(X_a,X_b)$
and $\partial_y F_{a}(1\!-\!X_a,1\!-\!X_b) =
({\langle q_a\rangle}/{\langle q_b\rangle}) \partial_a
\Psi_b(X_a,X_b)$,
gives
\begin{equation}\label{del_Ha}
\partial_u H_{a}(1,1) =
\frac{  R_a[1- \partial_b \Psi_{b}(X_a,X_b)]}
{\det[{\bf J}-{\bf I}]}.
\end{equation}
From Eq. (\ref{hybrid_condition_asym}) we see immediately that this
diverges at the critical point, meaning that
the mean size of avalanches triggered by random removal of vertices diverges
exactly at the point of the hybrid transition.
The mean
size of the avalanches can be related to the susceptibility,
of the  giant viable cluster to random damage, 
similar to the
susceptibility for ordinary percolation \cite{Stauffer1992}. 

Due to the similarity of Eq. (\ref{squareroot}) to the $k$-core
version \cite{Dorogovtsev2006b}, we can
expect that the size distribution of avalanches triggered by randomly
removed vertices obeys power law $p(s) \propto s^{-\sigma}$ with
$\sigma = 3/2$.

{\it Scale-free Networks}.---In ordinary percolation, and
even the $k$-core, networks with degree distributions that are asymptotically
powerlaws $P(q) \sim q^{-\gamma}$ may exhibit qualitatively different
transitions, especially when $\gamma < 3$.
To investigate such effects in the giant viable cluster, we
consider consider two uncorrelated scale-free networks, so $P(q_a,q_b) =
P_a(q_a)P_b(q_b)$, having 
powerlaw degree distributions with fixed minimum degree $q_0
= 1$ (then $\langle q\rangle \approx (\gamma-1)q_0/(\gamma-2)$), so that
$P_s(q_s) = \zeta(\gamma_s) q^{-\gamma_s}$
where $s$ takes the values $a$ or $b$.
As a control parameter we apply random damage to the system
as a whole so that vertices survive with probability $p$.
First consider the case that $\gamma_a < 3$ or $\gamma_b < 3$ (or
both).

The giant viable cluster is
necessarily a subgraph of the overlap between the giant-components of
each graph. We know from ordinary percolation that for $\gamma > 3$,
the giant component appears at a finite value of $p$
\cite{Cohen2002}. It follows that the giant viable cluster, also, cannot
appear from $p=0$; there must be a finite threshold $p_c$, (with a
hybrid transition)
 This is true even if one of
the networks has $\gamma_s < 3$.

The more interesting case is when 
$\gamma_a,\gamma_b<3$,
when the percolation threshold is zero for each
network when considered separately.
Let us write $\gamma_a = 2 + \delta_a$ and $\gamma_b =
2+\delta_b$, and examine the behavior for small $\delta_a$ and
$\delta_b$.
We proceed by assuming that in this situation, for $p$ near $p_c$,
Eqs. (\ref{Psi_s}) % and (\ref{Psi_b})
have a solution with small $X_a$, $X_b \ll 1$. Writing only
leading orders, 
\begin{equation}\label{Psi_a_sf}
\Psi_a(X_a,X_b) = p\frac{\pi^2}{6\, \delta_b}
X_a^{\delta_a}\left(X_b - X_b^{1+\delta_b}\right)
\end{equation}
and similarly for $\Psi_b(X_a,X_b)$.
The location of the critical point is found from
Eq. (\ref{hybrid_condition_asym}) which becomes
\begin{equation}\label{jump_asym_sf}
\delta_a + \delta_b = p \frac{\pi^2}{6} X_a^{\delta_a} X_b^{\delta_b}
\left(\frac{X_a}{X_b} + \frac{X_b}{X_a} \right)\,.
\end{equation}
Substituting Eq. (\ref{Psi_a_sf}) into Eq. (\ref{Psi_s}) and solving
with Eq. (\ref{jump_asym_sf}) we find 
 $X_s$ and $\mathcal{S}$ at $p_c$.
We find in general that the hybrid transition persists 
for $\delta_a, \delta_b \neq 0$, 
though the height of the
discontinuity at the hybrid
transition becomes extremely small for $\delta$ small.
In experiments or simulations, this could be misinterpreted as
evidence of a continuous phase transition.
We describe two representative cases. First, where $\delta_a \ll
\delta_b$, that is, $\gamma_a \to 2$ while $\gamma_b > 2$. 
We find that 
$p_c
\approx 1.19 \delta_b$, and the size of the giant viable cluster at
$p_c$ 
is $S_c
= A e^{-B/\delta_b}$ with $A \approx 3.36$ and $B \approx 2.89$.
We see that a hybrid transition occurs, albeit with an extremely small
discontinuity, at a
non-zero threshold $p_c$ as long as at least one of $\delta_a$ and
$\delta_b$ is not equal to zero. To examine the case that both tend to
zero, we consider the symmetric case $\delta_a = \delta_b \equiv
\delta$. Then $X_a = X_b \equiv X$, and
the discontinuity is found by requiring $\Psi'(X) = 1$
[from Eq. (\ref{hybrid_condition_asym})].
We find that $X_c = (1/2)^{1/\delta}$,
$p_c = 24 \delta / \pi^2$, and, $\mathcal{S}_c = {4^{1-1/\delta}}$.
The 
critical point, $p_c$, tends to $0$
as $\delta \to 0$, and 
$\mathcal{S}_c$
becomes very small even
for nonzero $\delta$, but vanishes completely as $\delta\to 0$.
See Fig.~\ref{S_vs_p}.
Expanding $\Psi(X)$ about $X_c$ we find 
again square-root scaling,
$X/X_c - 1 = A(p/p_c -1)^{1/2}$ with $A = 12/\pi^2 \delta p_c$,
which holds while $p-p_c \ll \delta^3$.

{\it Summary}.---We have given an algorithm for identifying the viable
clusters in any multiplex network. Under increasing damage, the giant
viable cluster
collapses in a discontinuous hybrid transition, in contrast to the
smooth continuous transition found in simplex networks. We have shown
that this transition is signaled by avalanches 
triggered by removing vertices at random. The mean size of the avalanches
diverges
as the collapse approaches. To understand this critical behavior,
which occurs only above the transition, we successfully identified
clusters of critical vertices.
 These clusters determine the
structure and statistics of avalanches of damage.
Avalanches sweep through the critical clusters
in a directed fashion, and it is the diverging size of these clusters 
which accounts for the criticality. This directed nature stands in
contrast to, for example, the corona clusters found in the $k$-core
problem \cite{Dorogovtsev2006b}.
Each critical cluster depends upon a keystone vertex
whose removal completely destroys the critical cluster.
These keystone vertices are good candidates for targeted
attack or immunization against such attacks.

\begin{acknowledgments}
This work was partially supported by FET IP Project
MULTIPLEX 317532 and by the PTDC projects SAU-
NEU/103904/2008, MAT/114515/2009, and PEst-C/
CTM/LA0025/2011, and post-doctoral fellowship
SFRH/BPD/74040/2010.
\end{acknowledgments}

\end{document}